\documentclass[aps,prl,twocolumn,showpacs,superscriptaddress,groupedaddress]{revtex4}  
\usepackage{graphicx}  
\usepackage{dcolumn}   
\usepackage{bm}        
\usepackage{amssymb}   

\begin{document}


\title{Aerodynamically-Optimal Hovering Is Not Stable}
\title{Optimal Hovering Conditions Are Not Stable}
\title{Stable Hovering is Not Aerodynamically-Optimal}
\title{Stability versus Maneuverability in Hovering Flight}

\author{Yangyang Huang}
\affiliation{Department of Aerospace and Mechanical Engineering, \\University of Southern California, Los Angeles, CA 90089, USA}
\author{Monika Nitsche}
\affiliation{Department of Mathematics and Statistics, \\University of New Mexico, Albuquerque, NM 87131, USA}
\author{Eva Kanso} \thanks{corresponding author: kanso@usc.edu}
\affiliation{Department of Aerospace and Mechanical Engineering, \\University of Southern California, Los Angeles, CA 90089, USA}

\date{\today}

\begin{abstract}
Insects and birds are often faced by opposing requirements for agile and stable flight. Here, we explore the interplay between aerodynamic effort,  maneuverability, and stability in a model system that consists of a $\Lambda$-shaped flyer hovering in a vertically oscillating airflow. We determine effective conditions that lead to periodic hovering in terms of two parameters: the flyer's shape (opening angle) and the effort (flow acceleration) needed to keep the flyer aloft. We find optimal shapes that minimize effort. We then examine hovering stability and observe a transition from unstable, yet maneuverable, to stable hovering. Interestingly, this transition occurs at post-optimal shapes, that is, at increased aerodynamic effort.  These results have profound implications on the interplay between stability and maneuverability in live organisms as well as on the design of man-made air vehicles.
\end{abstract}

\maketitle


The unsteady flow-structure interactions in flapping wing motions produce lift and thrust forces that allow insects and birds to fly forward or hover in place. The mechanisms responsible for the generation of these aerodynamic forces received a great deal of attention in recent experimental \cite{Ellington1996, Dickinson1999, Spedding2003, Birch2003, Thomas2004, Douglas2005} and theoretical \cite{Ramamurti2002, Minotti2002, Sun2002, Sun2004, Wang2000PRL, Wang2000JFM, Wang2004, Wang2007, Spagnolie2009} studies, mostly emphasizing the importance of leading-edge and wake vorticity in force production \cite{Sane2003, Wang2005}. 
However, the stability of flapping flight in response to environmental disturbances is less well explored \cite{Sun2014}.
Recent studies report conflicting accounts of intrinsic instability \cite{Sun2005, Sun2007, Faruque2010, Wu2012} and passive stability \cite{Taylor2002, Taylor2005, Ristroph2013}.

Live organisms certainly employ active feedback control during flight \cite{Ristroph2010, Gillies2011}, but it is not clear to what extent. Active stabilization requires additional \emph{effort} and thus energy expenditure. One can thus argue that passive stability reduces the effort required for flying. In this sense, it seems reasonable to conjecture that, from an evolutionary perspective, passive stability may have a positive selection value. However, stability can be thought of as ``resistance to change" which conflicts with maneuverability \cite{Taylor2002, Dudley2002}.  Unlike stability, there is no clear quantitative definition of maneuverability, which  we consider here to simply mean lack of stability.
Stable motions require extra effort to change, which could make sudden maneuvers energetically costly, whereas an unstable motion only needs a slight perturbation to change because the aerodynamic forces help  in moving the system away from its current state, making it easier to maneuver. Basically, there is a tradeoff between the effort required to maintain an unstable motion and that of causing a stable motion to change -- that is to say, a trade-off between stability and maneuverability.

\begin{figure}[!h]
\includegraphics[scale=1]{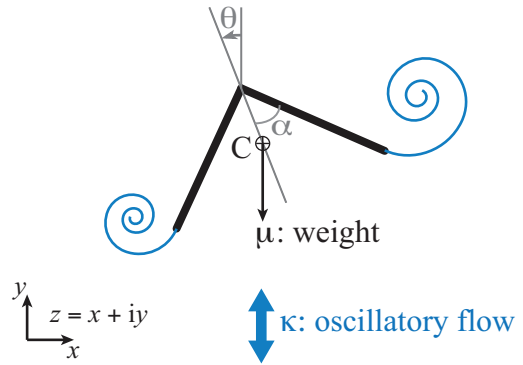}
\caption{A $\Lambda$-shaped flyer subject to gravity in a vertically-oscillating background flow.
\label{fig:schematic}}
\end{figure}

Whereas an assessment of the passive stability of live organisms is not feasible experimentally, an ingenious model system of an inanimate flyer was proposed recently as a proxy to flapping flight~\cite{Childress2006, Weathers2010, Liu2012}.  The experimental model consists of an upward-pointing pyramid-shaped object in a vertically oscillating airflow~\cite{Weathers2010, Liu2012}. The inanimate flyer generates sufficient aerodynamic force to keep aloft  and maintains balance passively during free flight. In~\cite{Liu2012}, the authors use clever arguments and simplifying approximations founded on a deep understanding of aerodynamics to obtain ``educated guesses" of the stabilizing mechanism without ever solving the coupled flow-structure interactions. In this work, we formulate a two-dimensional model of a $\Lambda$-shaped object  in an oscillating uniform flow, see Fig.~\ref{fig:schematic}.  This formulation enables us to quantitatively examine the aerodynamic forces required to keep the flyer aloft and the stabilizing aerodynamic moments. Most importantly, it provides a quantitative framework for exploring aerodynamic-optimal hovering conditions and for rigorously studying the transition from unstable, yet more maneuverable, to stable hovering. 

\begin{figure*}[!t]
\includegraphics[width=\linewidth]{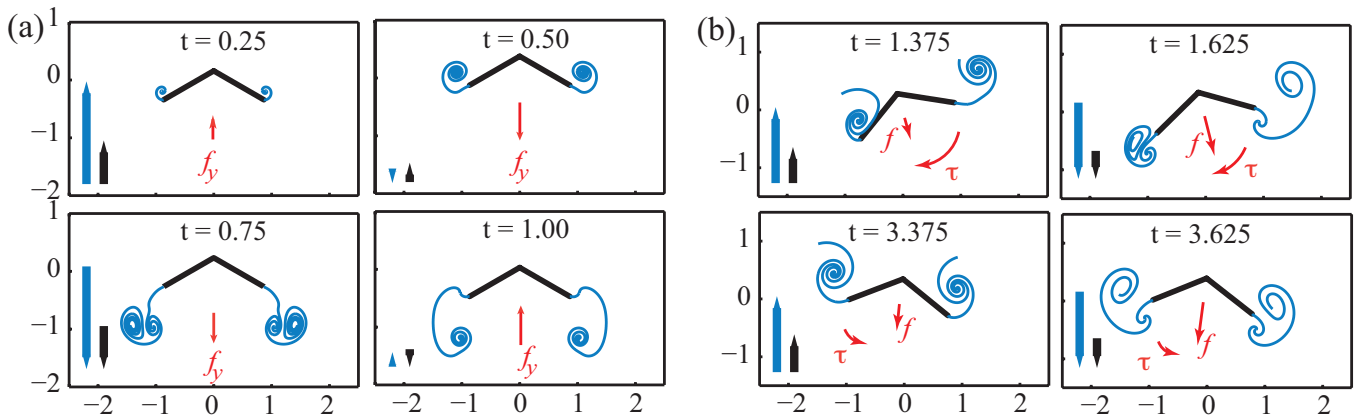}
\caption{$\Lambda$-flyer stably hovering in oscillatory flow: (a) Snapshots for vertically-upright initial conditions. The parameter values are set to $\alpha = 60^o$, $m = 8$, $\beta = 1$, $\kappa = 6.5$ and $T_{\rm diss} = 0.6$. Black and Blue arrows show the velocities of the flyer and the background flow respectively. Instantaneous aerodynamic force is shown in red arrow. (b) Snapshots when same flyer is initially tilted at an angle $\theta = 30^o$. Aerodynamic torque (shown in red) tends to restore the upright orientation of the flyer.} 
\label{fig:hoverexamples}
\end{figure*}

Our model $\Lambda$-flyer consists of two flat plates, of equal length $l$ and total mass $M$, joined at the apex at an angle $2 \alpha$, see Fig.~\ref{fig:schematic}. 
The background fluid of density $\rho_f$ oscillates vertically with velocity $U =  A (\pi f) \mathrm{sin}(2\pi f \,t)$, where $f$ is the oscillation frequency and $A$ is the top-bottom oscillation amplitude. 
Four relevant dimensionless parameters can be constructed: the mass $m = M/\rho_f l^2$ and weight $\mu = mg/lf^2$ of the $\Lambda$-flyer, and the amplitude  $\beta = A/l$ and acceleration $\kappa = Af^2/g$ of the background flow oscillations. Note that the parameter $\kappa$ can be  interpreted as a measure of the \textit{effort} needed to keep the flyer aloft.

Let $z_c = x_c + iy_c$ denote the position of the mass center $C$ of the $\Lambda$-flyer   in the complex $z$-plane ($i=\sqrt{-1}$) and let $\theta$ denote  its  orientation from the upward vertical (Fig.~\ref{fig:schematic}).
The equations governing its free motion under the effects of gravitational and aerodynamic forces are
\begin{equation}
m\ddot{z}_c = f_x + \mathrm{i}(f_y - \mu), \quad I\ddot{\theta} = \tau,
\label{eq:eom}
\end{equation}
where $I = m(1 - \frac{3}{4}\cos^2(\alpha))/3$ is the dimensionless moment of inertia, $f_x$, $f_y$ and $\tau$ are the aerodynamic forces and torque.

We simulate the flow using a vortex sheet model in the inviscid fluid context. The $\Lambda$-flyer is modeled as a bound vortex sheet that satisfies zero normal flow through the flyer. A point vortex is released at each time step from the two outer edges of the $\Lambda$-flyer, and the shed vorticity is modeled as a regularized free sheet \cite{Krasny1986, Nitsche1994, Jones2003, Jones2005, Alben2008, Michelin2008}. No separation is allowed at the apex. 
Here, we follow the algorithm in \cite{Nitsche1994} for imposing the Kutta condition that determines the amount of circulation shed from the outer two edges at each time step. 
The vortex sheet model depends on the regularization parameter for the free sheet, which in the results below is set to $\delta/l = 0.1$. 
By way of validation, we confirmed that our numerical scheme gives identical results for examples presented in \cite{Jones2003,Jones2005} of 
driven flat plates  and plates freely falling under gravity, even though the implementation details differ significantly. 
Finally, to emulate the effect of viscosity, we allow the shed vortex sheet to decay gradually by dissipating each incremental point vortex after a finite time $T_{{\rm diss}}$ from the time it is shed in the fluid. 
Larger $T_{\rm diss}$ implies  lower fluid viscosity.  A closed-form expression that rigorously links $T_{\rm diss}$ to the kinematic fluid viscosity $\nu$ is not readily available, however, using approximate arguments based on the Lamb-Oseen solution, we choose $T_{\rm diss}$ such that $\nu T_{\rm diss}$ is small, where $\nu$ is the normalized viscosity of air. 

\begin{figure}[!t]
\includegraphics[width=0.9\linewidth]{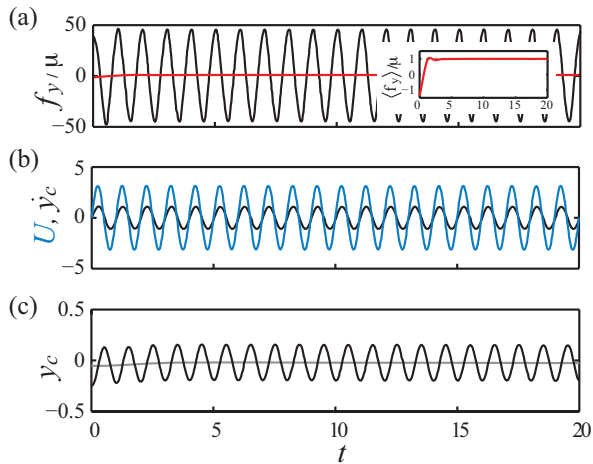}
\caption{Periodic hovering of the $\Lambda$-flyer shown in Fig.~\ref{fig:hoverexamples}(a) for vertically-upright initial conditions: (a) Aerodynamic force normalized by the weight $\mu$ oscillates with the background flow such that its time average reaches $1$ after a short transience ($\sim 3 T$).  
(b) Flyer's velocity $\dot{y}_c$ oscillates at the same frequency as the background flow $U$ but with smaller amplitude. (c) 
Flyer's vertical  position $y_c$ also oscillates but such that the change in the $T$-averaged position is zero. }
\label{fig:hoveringdata}
\end{figure}

\begin{figure}[!t]
\includegraphics[width=0.8\linewidth]{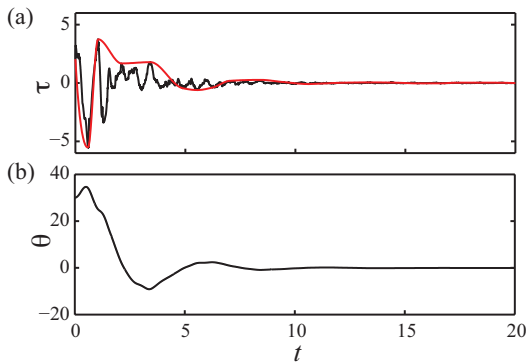}
\caption{Stable nonlinear response of the $\Lambda$-flyer shown in Fig.~\ref{fig:hoverexamples}(b) for tilted initial conditions:  (a) The envelope (red) of the aerodynamic torque (black)  fluctuates out of phase relative to the flyer's orientation $\theta$ shown in (b), thus producing a restorative effect that causes the flyer to recover the vertically-upright hovering. }
\label{fig:stabilitydata}
\end{figure}

We first examine the behavior of a flyer undergoing periodic hovering motion.
Fig.~\ref{fig:hoverexamples}(a) depicts snapshots of the hovering motion and vortical wake for a flyer with angle $\alpha = 60^o$ and mass $m = 8$ in an oscillating flow of amplitude $\beta=1$, acceleration $\kappa = 6.5$, and dissipation parameter $T_{\rm{diss}}=0.6$. The total simulation time is $t_{\rm end} = 20T$, where $T = 2\pi/f$ is the oscillation period of the background flow. The flyer is subject to zero initial velocity $\dot{x}_c(0) = \dot{y}_c(0) = 0$ and tilt conditions $\theta(0) = \dot{\theta}(0) =0$. Clearly, during the up-flow,  vortices are generated at the two outer edges of the flyer. These vortices combine with the vortices generated during the down-flow to form two vortex dipoles that initially move vertically down. This downwash results in a lift force that balances the weight of the flyer keeping it aloft, as noted qualitatively in \cite{Liu2012} . 
Quantitatively,  the  aerodynamic torque $\tau$ and horizontal force $f_x$ acting on the flyer  are identically zero as expected from symmetry considerations while the vertical force $f_y$ oscillates periodically from positive to negative at the same frequency as the background flow such that its $T$-averaged value $\langle f_y \rangle  = \frac{1}{T}\left(\int_t^{t+T} f_y(\tilde{t}) d\tilde{t}\right)$ when normalized by the flyer's weight $\mu$ is equal to $\langle f_y \rangle/\mu  =1$ (Fig~\ref{fig:hoveringdata}(a)). Vortex shedding is essential for the generation of these lift forces.
The flyer responds by oscillating up and down at speeds smaller than those of the background oscillatory flow (Fig~\ref{fig:hoveringdata}(b)) such that it hovers  around its initial vertical position (Fig~\ref{fig:hoveringdata}(c)). 
By hovering, we mean that the change $\Delta y_c =  \left. \langle y_c \rangle \right|_{t_{\rm end}} - \left.\langle y_c \rangle \right|_{0}$  in the $T$-averaged vertical position $\langle y_c \rangle = \frac{1}{T}\left(\int_t^{t+T} y_c(\tilde{t}) d\tilde{t}\right)$ is equal to zero. 

This hovering motion is stable to initial perturbation, which we impose here on the initial  tilt angle $\theta(0)$.
Surprisingly, the flyer recovers the upright orientation and continues to hover stably for a range of initial perturbations as large as $\theta(0) = 76^o$. For $\theta(0) = 30^o$, snapshots of the flyer and its wake during the recovery phase are depicted in Fig.~\ref{fig:hoverexamples}(b). When the flyer is tilted to one side, the left-right symmetry of the shed vorticity is broken, which leads to stronger vorticity shed sideways from the edge with the larger angle of attack. The sideward vorticity creates a restorative aerodynamic torque
as argued  qualitatively in \cite{Liu2012} and depicted quantitatively in Fig.~\ref{fig:stabilitydata}. Here, both the torque envelope (shown in red line) and the orientation of the flyer fluctuate out of phase relative to each other, indicating the restorative effect of the aerodynamic torque. The fluctuations decrease in amplitude and eventually approach zero as the flyer recovers its upright orientation.

\begin{figure*}
\includegraphics[width=\linewidth]{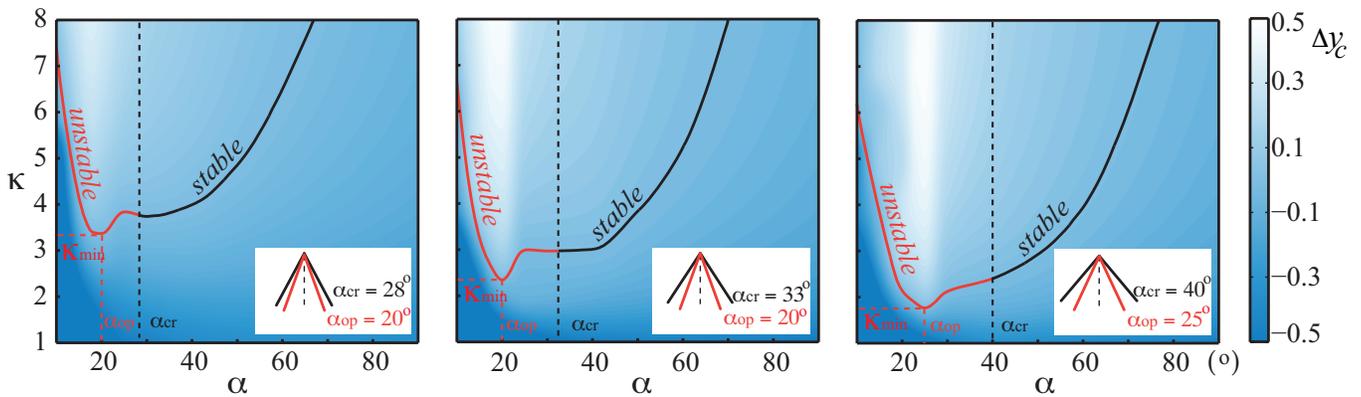}
\caption{Parameter space $(\alpha,\kappa)$: from left to right, fluid viscosity decreases as dissipation time increases $T_{\rm diss}=0.6$, $0.65$ and $0.7$. In each case, the \emph{hovering curve} corresponds to symmetric hovering ($\triangle y_c=0$). Above ($\triangle y_c>0$) and below ($\triangle y_c<0$) this line, the  flyer ascends and descends, respectively. Flyers of optimal shape $\alpha_{\rm op}$ require minimum flow accelerations $\kappa_{\rm min}$ for hovering.  A stability study of these hovering motion shows a transition from unstable to stable hovering as $\alpha$ increases. The transition occurs at $\alpha_{\rm cr} > \alpha_{\rm op}$.}
\label{fig:paramspace}
\end{figure*}

We now determine effective conditions for hovering as a function of two parameters: the flyer's shape described by the opening angle $\alpha$ and the \emph{effort} needed of the oscillating flow expressed by the flow acceleration parameter $\kappa$. We set $m =8$, $\beta = 1$ and we vary $\alpha$ from $10^o$ to $90^o$ and $\kappa$ from $1$ to $8$. Note that, for a flyer of a given shape, there is an associated effort or flow acceleration that keeps the flyer aloft when starting in its upright position with zero initial velocity. Stronger or weaker efforts would cause the flyer to ascend or descend. That is to say, each point in the parameter space $(\alpha, \kappa)$ represents one of three types of behavior: ascending $(\Delta y_c >0)$, hovering $(\Delta y_c =0)$ or descending $(\Delta y_c <0)$. The hovering condition $\Delta y_c =0$ defines a \emph{hovering curve} in the $(\alpha, \kappa)$-plane  as depicted in Fig.~\ref{fig:paramspace} for three cases: $T_{\rm diss} =0.6, 0.65$ and $0.7$, corresponding to decreasing fluid ``viscosity.''  In all three cases, there exists an optimal shape $\alpha_{\rm op}$ hovering curve admits a global minimum $\kappa_ {\rm min}$, that is, for which the effort $\kappa$ required to hover is minimum. The value of the minimum effort  $\kappa_{\rm min}$ decreases as the ``viscosity" decreases, which can be intuitively understood on the ground that, at lower viscosity, the shed vortices responsible for the lift production are longer lived.

We then analyze the passive stability of all points on the hovering curve by imposing a small initial perturbation $\theta(0)=1^o$ and solving the fully nonlinear governing equations of motion in~(\ref{eq:eom}). In particular, we focus on the time evolution of the tilt angle $\theta$:
 if it oscillates with decreasing or constant amplitude, we say the flyer is passively stable. If the amplitude of $\theta$ grows in time, the flyer is unstable. By mapping out these stability results to the hovering curves in Fig.~\ref{fig:paramspace}, we see a transition from unstable to stable hovering as the opening angle of the flyer increases. 
 Most importantly, the transition from unstable to stable hovering occurs at a critical shape $\alpha_\mathrm{cr}$ that is post-optimal ($\alpha_{\rm cr} > \alpha_{\rm op}$). This result leads to interesting insights on the interplay between maneuverability and stability in hovering flights.
 

$\Lambda$-flyers with optimal shapes $\alpha_{\rm op}$ operating at minimum aerodynamic effort $\kappa_{min}$ produce hovering motions that are passively unstable. One should therefore be careful when optimizing for aerodynamic effort alone without paying attention to motion stability. In so doing, one would obtain optimal flyers that, although more maneuverable, would require active stabilization mechanisms. Active stabilization requires aerodynamic effort that may be even larger than the effort required for passive stability. This interpretation assumes that, when evaluating or designing flyers, one should opt for either stability or maneuverability. However, the results in Fig.~\ref{fig:paramspace} lend themselves to a far richer explanation. They suggest that a $\Lambda$-flyer that could actively change its shape, as in the case of live organisms, can smoothly switch from passively stable to unstable, yet more maneuverable, states by decreasing its opening angle. They also suggest that, although passive stability is not free (it comes at a higher effort $\kappa$), switching from stable to maneuverable states requires no extra effort, it rather requires a decrease in the aerodynamic effort $\kappa$ because the transition $\alpha_{\rm cr}$ occurs post-optimally for $\alpha_{\rm cr} > \alpha_{\rm op}$. Accordingly, we conjecture that, to fulfill the two requirements of passive stability and maneuverability, a good design practice both in nature and in man-made aerial vehicles  is to position the stability limit at a post-optimal location in the parameter space. 

Future extensions of this work will include studying the effects of body deformation and body elasticity on the aerodynamic effort and stability of flapping flight, both in two- and three-dimensions.

\bibliography{Huang2014}

\begin{thebibliography}{37}
\expandafter\ifx\csname natexlab\endcsname\relax\def\natexlab#1{#1}\fi
\expandafter\ifx\csname bibnamefont\endcsname\relax
  \def\bibnamefont#1{#1}\fi
\expandafter\ifx\csname bibfnamefont\endcsname\relax
  \def\bibfnamefont#1{#1}\fi
\expandafter\ifx\csname citenamefont\endcsname\relax
  \def\citenamefont#1{#1}\fi
\expandafter\ifx\csname url\endcsname\relax
  \def\url#1{\texttt{#1}}\fi
\expandafter\ifx\csname urlprefix\endcsname\relax\def\urlprefix{URL }\fi
\providecommand{\bibinfo}[2]{#2}
\providecommand{\eprint}[2][]{\url{#2}}

\bibitem[{\citenamefont{Ellington et~al.}(1996)\citenamefont{Ellington, van~den
  Berg, Willmott, and Thomas}}]{Ellington1996}
\bibinfo{author}{\bibfnamefont{C.~P.} \bibnamefont{Ellington}},
  \bibinfo{author}{\bibfnamefont{C.}~\bibnamefont{van~den Berg}},
  \bibinfo{author}{\bibfnamefont{A.~P.} \bibnamefont{Willmott}},
  \bibnamefont{and} \bibinfo{author}{\bibfnamefont{A.~L.~R.}
  \bibnamefont{Thomas}}, \bibinfo{journal}{Nature}
  \textbf{\bibinfo{volume}{384}}, \bibinfo{pages}{626} (\bibinfo{year}{1996}).

\bibitem[{\citenamefont{Dickinson et~al.}(1999)\citenamefont{Dickinson,
  Lehmann, and Sane}}]{Dickinson1999}
\bibinfo{author}{\bibfnamefont{M.~H.} \bibnamefont{Dickinson}},
  \bibinfo{author}{\bibfnamefont{F.-O.} \bibnamefont{Lehmann}},
  \bibnamefont{and} \bibinfo{author}{\bibfnamefont{S.~P.} \bibnamefont{Sane}},
  \bibinfo{journal}{Science} \textbf{\bibinfo{volume}{284}},
  \bibinfo{pages}{1954} (\bibinfo{year}{1999}).

\bibitem[{\citenamefont{Spedding et~al.}(2003)\citenamefont{Spedding,
  Ros{\'e}n, and Hedenstr{\"o}m}}]{Spedding2003}
\bibinfo{author}{\bibfnamefont{G.}~\bibnamefont{Spedding}},
  \bibinfo{author}{\bibfnamefont{M.}~\bibnamefont{Ros{\'e}n}},
  \bibnamefont{and}
  \bibinfo{author}{\bibfnamefont{A.}~\bibnamefont{Hedenstr{\"o}m}},
  \bibinfo{journal}{J. Exp. Biol.} \textbf{\bibinfo{volume}{206}},
  \bibinfo{pages}{2313} (\bibinfo{year}{2003}).

\bibitem[{\citenamefont{Birch and Dickinson}(2003)}]{Birch2003}
\bibinfo{author}{\bibfnamefont{J.~M.} \bibnamefont{Birch}} \bibnamefont{and}
  \bibinfo{author}{\bibfnamefont{M.~H.} \bibnamefont{Dickinson}},
  \bibinfo{journal}{J. Exp. Biol.} \textbf{\bibinfo{volume}{206}},
  \bibinfo{pages}{2257} (\bibinfo{year}{2003}).

\bibitem[{\citenamefont{Thomas et~al.}(2004)\citenamefont{Thomas, Taylor,
  Srygley, Nudds, and Bomphrey}}]{Thomas2004}
\bibinfo{author}{\bibfnamefont{A.~L.~R.} \bibnamefont{Thomas}},
  \bibinfo{author}{\bibfnamefont{G.~K.} \bibnamefont{Taylor}},
  \bibinfo{author}{\bibfnamefont{R.~B.} \bibnamefont{Srygley}},
  \bibinfo{author}{\bibfnamefont{R.~L.} \bibnamefont{Nudds}}, \bibnamefont{and}
  \bibinfo{author}{\bibfnamefont{R.~J.} \bibnamefont{Bomphrey}},
  \bibinfo{journal}{J. Exp. Biol.} \textbf{\bibinfo{volume}{207}},
  \bibinfo{pages}{4299} (\bibinfo{year}{2004}).

\bibitem[{\citenamefont{Warrick et~al.}(2005)\citenamefont{Warrick, Tobalske,
  and Powers}}]{Douglas2005}
\bibinfo{author}{\bibfnamefont{D.~R.} \bibnamefont{Warrick}},
  \bibinfo{author}{\bibfnamefont{B.~W.} \bibnamefont{Tobalske}},
  \bibnamefont{and} \bibinfo{author}{\bibfnamefont{D.~R.}
  \bibnamefont{Powers}}, \bibinfo{journal}{Nature}
  \textbf{\bibinfo{volume}{435}}, \bibinfo{pages}{1094} (\bibinfo{year}{2005}).

\bibitem[{\citenamefont{Ramamurti and Sandberg}(2002)}]{Ramamurti2002}
\bibinfo{author}{\bibfnamefont{R.}~\bibnamefont{Ramamurti}} \bibnamefont{and}
  \bibinfo{author}{\bibfnamefont{W.~C.} \bibnamefont{Sandberg}},
  \bibinfo{journal}{J. Exp. Biol.} \textbf{\bibinfo{volume}{205}},
  \bibinfo{pages}{1507} (\bibinfo{year}{2002}).

\bibitem[{\citenamefont{Minotti}(2002)}]{Minotti2002}
\bibinfo{author}{\bibfnamefont{F.~O.} \bibnamefont{Minotti}},
  \bibinfo{journal}{Phys. Rev. E} \textbf{\bibinfo{volume}{66}},
  \bibinfo{pages}{051907} (\bibinfo{year}{2002}).

\bibitem[{\citenamefont{Sun and Tang}(2002)}]{Sun2002}
\bibinfo{author}{\bibfnamefont{M.}~\bibnamefont{Sun}} \bibnamefont{and}
  \bibinfo{author}{\bibfnamefont{J.}~\bibnamefont{Tang}}, \bibinfo{journal}{J.
  Exp. Biol.} \textbf{\bibinfo{volume}{205}}, \bibinfo{pages}{55}
  (\bibinfo{year}{2002}).

\bibitem[{\citenamefont{Sun and Lan}(2004)}]{Sun2004}
\bibinfo{author}{\bibfnamefont{M.}~\bibnamefont{Sun}} \bibnamefont{and}
  \bibinfo{author}{\bibfnamefont{S.~L.} \bibnamefont{Lan}},
  \bibinfo{journal}{J. Exp. Biol.} \textbf{\bibinfo{volume}{207}},
  \bibinfo{pages}{1887} (\bibinfo{year}{2004}).

\bibitem[{\citenamefont{Wang}(2000{\natexlab{a}})}]{Wang2000PRL}
\bibinfo{author}{\bibfnamefont{Z.~J.} \bibnamefont{Wang}},
  \bibinfo{journal}{Phys. Rev. Lett.} \textbf{\bibinfo{volume}{85}},
  \bibinfo{pages}{2216} (\bibinfo{year}{2000}{\natexlab{a}}).

\bibitem[{\citenamefont{Wang}(2000{\natexlab{b}})}]{Wang2000JFM}
\bibinfo{author}{\bibfnamefont{Z.~J.} \bibnamefont{Wang}}, \bibinfo{journal}{J.
  Fluid Mech.} \textbf{\bibinfo{volume}{410}}, \bibinfo{pages}{323}
  (\bibinfo{year}{2000}{\natexlab{b}}).

\bibitem[{\citenamefont{Wang et~al.}(2004)\citenamefont{Wang, Birch, and
  Dickinson}}]{Wang2004}
\bibinfo{author}{\bibfnamefont{Z.~J.} \bibnamefont{Wang}},
  \bibinfo{author}{\bibfnamefont{J.~M.} \bibnamefont{Birch}}, \bibnamefont{and}
  \bibinfo{author}{\bibfnamefont{M.~H.} \bibnamefont{Dickinson}},
  \bibinfo{journal}{J. Exp. Biol.} \textbf{\bibinfo{volume}{207}},
  \bibinfo{pages}{449} (\bibinfo{year}{2004}).

\bibitem[{\citenamefont{Wang and Russell}(2007)}]{Wang2007}
\bibinfo{author}{\bibfnamefont{Z.~J.} \bibnamefont{Wang}} \bibnamefont{and}
  \bibinfo{author}{\bibfnamefont{D.}~\bibnamefont{Russell}},
  \bibinfo{journal}{Phys. Rev. Lett.} \textbf{\bibinfo{volume}{99}},
  \bibinfo{pages}{148101} (\bibinfo{year}{2007}).

\bibitem[{\citenamefont{Spagnolie and Shelley}(2009)}]{Spagnolie2009}
\bibinfo{author}{\bibfnamefont{S.~E.} \bibnamefont{Spagnolie}}
  \bibnamefont{and} \bibinfo{author}{\bibfnamefont{M.~J.}
  \bibnamefont{Shelley}}, \bibinfo{journal}{Phys. Fluids}
  \textbf{\bibinfo{volume}{21}}, \bibinfo{eid}{013103} (\bibinfo{year}{2009}).

\bibitem[{\citenamefont{Sane}(2003)}]{Sane2003}
\bibinfo{author}{\bibfnamefont{S.~P.} \bibnamefont{Sane}}, \bibinfo{journal}{J.
  Exp. Biol.} \textbf{\bibinfo{volume}{206}}, \bibinfo{pages}{4191}
  (\bibinfo{year}{2003}).

\bibitem[{\citenamefont{Wang}(2005)}]{Wang2005}
\bibinfo{author}{\bibfnamefont{Z.~J.} \bibnamefont{Wang}},
  \bibinfo{journal}{Annu. Rev. Fluid Mech.} \textbf{\bibinfo{volume}{37}},
  \bibinfo{pages}{183} (\bibinfo{year}{2005}).

\bibitem[{\citenamefont{Sun}(2014)}]{Sun2014}
\bibinfo{author}{\bibfnamefont{M.}~\bibnamefont{Sun}}, \bibinfo{journal}{Rev.
  Mod. Phys.} \textbf{\bibinfo{volume}{86}}, \bibinfo{pages}{615}
  (\bibinfo{year}{2014}).

\bibitem[{\citenamefont{Sun and Xiong}(2005)}]{Sun2005}
\bibinfo{author}{\bibfnamefont{M.}~\bibnamefont{Sun}} \bibnamefont{and}
  \bibinfo{author}{\bibfnamefont{Y.}~\bibnamefont{Xiong}}, \bibinfo{journal}{J.
  Exp. Biol.} \textbf{\bibinfo{volume}{208}}, \bibinfo{pages}{447}
  (\bibinfo{year}{2005}).

\bibitem[{\citenamefont{Sun et~al.}(2007)\citenamefont{Sun, Wang, and
  Xiong}}]{Sun2007}
\bibinfo{author}{\bibfnamefont{M.}~\bibnamefont{Sun}},
  \bibinfo{author}{\bibfnamefont{J.}~\bibnamefont{Wang}}, \bibnamefont{and}
  \bibinfo{author}{\bibfnamefont{Y.}~\bibnamefont{Xiong}},
  \bibinfo{journal}{Acta Mech. Sin.} \textbf{\bibinfo{volume}{23}},
  \bibinfo{pages}{231} (\bibinfo{year}{2007}).

\bibitem[{\citenamefont{Faruque and Humbert}(2010)}]{Faruque2010}
\bibinfo{author}{\bibfnamefont{I.}~\bibnamefont{Faruque}} \bibnamefont{and}
  \bibinfo{author}{\bibfnamefont{J.~S.} \bibnamefont{Humbert}},
  \bibinfo{journal}{J. Theor. Biol.} \textbf{\bibinfo{volume}{264}},
  \bibinfo{pages}{538 } (\bibinfo{year}{2010}).

\bibitem[{\citenamefont{Wu and Sun}(2012)}]{Wu2012}
\bibinfo{author}{\bibfnamefont{J.~H.} \bibnamefont{Wu}} \bibnamefont{and}
  \bibinfo{author}{\bibfnamefont{M.}~\bibnamefont{Sun}}, \bibinfo{journal}{J.
  R. Soc. Interface} \textbf{\bibinfo{volume}{9}}, \bibinfo{pages}{2033}
  (\bibinfo{year}{2012}).

\bibitem[{\citenamefont{Taylor and Thomas}(2002)}]{Taylor2002}
\bibinfo{author}{\bibfnamefont{G.~K.} \bibnamefont{Taylor}} \bibnamefont{and}
  \bibinfo{author}{\bibfnamefont{A.~L.~R.} \bibnamefont{Thomas}},
  \bibinfo{journal}{J. Theor. Biol.} \textbf{\bibinfo{volume}{214}},
  \bibinfo{pages}{351} (\bibinfo{year}{2002}).

\bibitem[{\citenamefont{Taylor and {\.Z}bikowski}(2005)}]{Taylor2005}
\bibinfo{author}{\bibfnamefont{G.~K.} \bibnamefont{Taylor}} \bibnamefont{and}
  \bibinfo{author}{\bibfnamefont{R.}~\bibnamefont{{\.Z}bikowski}},
  \bibinfo{journal}{J. R. Soc. Interface} \textbf{\bibinfo{volume}{2}},
  \bibinfo{pages}{197} (\bibinfo{year}{2005}).

\bibitem[{\citenamefont{Ristroph et~al.}(2013)\citenamefont{Ristroph, Ristroph,
  Morozova, Bergou, Chang, Guckenheimer, Wang, and Cohen}}]{Ristroph2013}
\bibinfo{author}{\bibfnamefont{L.}~\bibnamefont{Ristroph}},
  \bibinfo{author}{\bibfnamefont{G.}~\bibnamefont{Ristroph}},
  \bibinfo{author}{\bibfnamefont{S.}~\bibnamefont{Morozova}},
  \bibinfo{author}{\bibfnamefont{A.~J.} \bibnamefont{Bergou}},
  \bibinfo{author}{\bibfnamefont{S.}~\bibnamefont{Chang}},
  \bibinfo{author}{\bibfnamefont{J.}~\bibnamefont{Guckenheimer}},
  \bibinfo{author}{\bibfnamefont{Z.~J.} \bibnamefont{Wang}}, \bibnamefont{and}
  \bibinfo{author}{\bibfnamefont{I.}~\bibnamefont{Cohen}}, \bibinfo{journal}{J.
  R. Soc. Interface} \textbf{\bibinfo{volume}{10}}, \bibinfo{pages}{20130237}
  (\bibinfo{year}{2013}).

\bibitem[{\citenamefont{Ristroph et~al.}(2010)\citenamefont{Ristroph, Bergou,
  Ristroph, Coumes, Berman, Guckenheimer, Wang, and Cohen}}]{Ristroph2010}
\bibinfo{author}{\bibfnamefont{L.}~\bibnamefont{Ristroph}},
  \bibinfo{author}{\bibfnamefont{A.~J.} \bibnamefont{Bergou}},
  \bibinfo{author}{\bibfnamefont{G.}~\bibnamefont{Ristroph}},
  \bibinfo{author}{\bibfnamefont{K.}~\bibnamefont{Coumes}},
  \bibinfo{author}{\bibfnamefont{G.~J.} \bibnamefont{Berman}},
  \bibinfo{author}{\bibfnamefont{J.}~\bibnamefont{Guckenheimer}},
  \bibinfo{author}{\bibfnamefont{Z.~J.} \bibnamefont{Wang}}, \bibnamefont{and}
  \bibinfo{author}{\bibfnamefont{I.}~\bibnamefont{Cohen}},
  \bibinfo{journal}{Proc. Natl. Acad. Sci. U.S.A.}
  \textbf{\bibinfo{volume}{107}}, \bibinfo{pages}{4820} (\bibinfo{year}{2010}).

\bibitem[{\citenamefont{Gillies et~al.}(2011)\citenamefont{Gillies, Thomas, and
  Taylor}}]{Gillies2011}
\bibinfo{author}{\bibfnamefont{J.~A.} \bibnamefont{Gillies}},
  \bibinfo{author}{\bibfnamefont{A.~L.~R.} \bibnamefont{Thomas}},
  \bibnamefont{and} \bibinfo{author}{\bibfnamefont{G.~K.}
  \bibnamefont{Taylor}}, \bibinfo{journal}{J. Avian Biol.}
  \textbf{\bibinfo{volume}{42}}, \bibinfo{pages}{377} (\bibinfo{year}{2011}).

\bibitem[{\citenamefont{Dudley}(2002)}]{Dudley2002}
\bibinfo{author}{\bibfnamefont{R.}~\bibnamefont{Dudley}},
  \bibinfo{journal}{Integr. Comp. Biol.} \textbf{\bibinfo{volume}{42}},
  \bibinfo{pages}{135} (\bibinfo{year}{2002}).

\bibitem[{\citenamefont{Childress et~al.}(2006)\citenamefont{Childress,
  Vandenberghe, and Zhang}}]{Childress2006}
\bibinfo{author}{\bibfnamefont{S.}~\bibnamefont{Childress}},
  \bibinfo{author}{\bibfnamefont{N.}~\bibnamefont{Vandenberghe}},
  \bibnamefont{and} \bibinfo{author}{\bibfnamefont{J.}~\bibnamefont{Zhang}},
  \bibinfo{journal}{Phys. Fluids} \textbf{\bibinfo{volume}{18}},
  \bibinfo{eid}{117103} (\bibinfo{year}{2006}).

\bibitem[{\citenamefont{Weathers et~al.}(2010)\citenamefont{Weathers, Folie,
  Liu, Childress, and Zhang}}]{Weathers2010}
\bibinfo{author}{\bibfnamefont{A.}~\bibnamefont{Weathers}},
  \bibinfo{author}{\bibfnamefont{B.}~\bibnamefont{Folie}},
  \bibinfo{author}{\bibfnamefont{B.}~\bibnamefont{Liu}},
  \bibinfo{author}{\bibfnamefont{S.}~\bibnamefont{Childress}},
  \bibnamefont{and} \bibinfo{author}{\bibfnamefont{J.}~\bibnamefont{Zhang}},
  \bibinfo{journal}{J. Fluid Mech.} \textbf{\bibinfo{volume}{650}},
  \bibinfo{pages}{415} (\bibinfo{year}{2010}).

\bibitem[{\citenamefont{Liu et~al.}(2012)\citenamefont{Liu, Ristroph, Weathers,
  Childress, and Zhang}}]{Liu2012}
\bibinfo{author}{\bibfnamefont{B.}~\bibnamefont{Liu}},
  \bibinfo{author}{\bibfnamefont{L.}~\bibnamefont{Ristroph}},
  \bibinfo{author}{\bibfnamefont{A.}~\bibnamefont{Weathers}},
  \bibinfo{author}{\bibfnamefont{S.}~\bibnamefont{Childress}},
  \bibnamefont{and} \bibinfo{author}{\bibfnamefont{J.}~\bibnamefont{Zhang}},
  \bibinfo{journal}{Phys. Rev. Lett.} \textbf{\bibinfo{volume}{108}},
  \bibinfo{pages}{068103} (\bibinfo{year}{2012}).

\bibitem[{\citenamefont{Krasny}(1986)}]{Krasny1986}
\bibinfo{author}{\bibfnamefont{R.}~\bibnamefont{Krasny}}, \bibinfo{journal}{J.
  Comput. Phys.} \textbf{\bibinfo{volume}{65}}, \bibinfo{pages}{292 }
  (\bibinfo{year}{1986}).

\bibitem[{\citenamefont{Nitsche and Krasny}(1994)}]{Nitsche1994}
\bibinfo{author}{\bibfnamefont{M.}~\bibnamefont{Nitsche}} \bibnamefont{and}
  \bibinfo{author}{\bibfnamefont{R.}~\bibnamefont{Krasny}},
  \bibinfo{journal}{J. Fluid Mech.} \textbf{\bibinfo{volume}{276}},
  \bibinfo{pages}{139} (\bibinfo{year}{1994}).

\bibitem[{\citenamefont{Jones}(2003)}]{Jones2003}
\bibinfo{author}{\bibfnamefont{M.~A.} \bibnamefont{Jones}},
  \bibinfo{journal}{J. Fluid Mech.} \textbf{\bibinfo{volume}{496}},
  \bibinfo{pages}{405} (\bibinfo{year}{2003}).

\bibitem[{\citenamefont{Jones and Shelley}(2005)}]{Jones2005}
\bibinfo{author}{\bibfnamefont{M.~A.} \bibnamefont{Jones}} \bibnamefont{and}
  \bibinfo{author}{\bibfnamefont{M.~J.} \bibnamefont{Shelley}},
  \bibinfo{journal}{J. Fluid Mech.} \textbf{\bibinfo{volume}{540}},
  \bibinfo{pages}{393} (\bibinfo{year}{2005}).

\bibitem[{\citenamefont{Alben and Shelley}(2008)}]{Alben2008}
\bibinfo{author}{\bibfnamefont{S.}~\bibnamefont{Alben}} \bibnamefont{and}
  \bibinfo{author}{\bibfnamefont{M.~J.} \bibnamefont{Shelley}},
  \bibinfo{journal}{Phys. Rev. Lett.} \textbf{\bibinfo{volume}{100}},
  \bibinfo{pages}{074301} (\bibinfo{year}{2008}).

\bibitem[{\citenamefont{Michelin et~al.}(2008)\citenamefont{Michelin,
  Llewellyn~Smith, and Glover}}]{Michelin2008}
\bibinfo{author}{\bibfnamefont{S.}~\bibnamefont{Michelin}},
  \bibinfo{author}{\bibfnamefont{S.~G.} \bibnamefont{Llewellyn~Smith}},
  \bibnamefont{and} \bibinfo{author}{\bibfnamefont{B.~J.}
  \bibnamefont{Glover}}, \bibinfo{journal}{J. Fluid Mech.}
  \textbf{\bibinfo{volume}{617}}, \bibinfo{pages}{1} (\bibinfo{year}{2008}).

\end{thebibliography}

\end{document}